\documentclass[fleqn,usenatbib]{mnras}

\usepackage{newtxtext,newtxmath}
\usepackage[utf8]{inputenc}
\usepackage[T1]{fontenc}
\usepackage{ae,aecompl}

\usepackage{graphicx}	% Including figure files
\usepackage{amsmath}	% Advanced maths commands
\usepackage{booktabs}

\usepackage[dvipsnames]{xcolor}

\newcommand{\lf}[1]{{\bf\color{Orchid}#1}}
\title[ML Approach for Seeing Correction in Flares]{A Machine Learning Approach to Correcting Atmospheric Seeing in Solar Flare Observations}

\author[J. A. Armstrong et al.]{
John A. Armstrong,$^{1}$\thanks{E-mail: j.armstrong.2@research.gla.ac.uk}
Lyndsay Fletcher$^{1, 2}$
\\
$^{1}$SUPA School of Physics and Astronomy, University of Glasgow, Glasogw, G12 8QQ, Scotland, U.K.\\
$^{2}$Rosseland Centre for Solar Physics, University of Oslo, P.O. Box 1029 Blindem, NO-0315 Oslo, Norway\\
}

\date{Accepted XXX. Received YYY; in original form ZZZ}

\pubyear{2020}

\hypersetup{draft}
\begin{document}
\label{firstpage}
\pagerange{\pageref{firstpage}--\pageref{lastpage}}
\maketitle

\begin{abstract}
Current post-processing techniques for the correction of atmospheric seeing in solar observations -- such as Speckle interferometry and Phase Diversity methods -- have limitations when it comes to their reconstructive capabilities of solar flare observations.
This, combined with the sporadic nature of flares meaning observers cannot wait until seeing conditions are optimal before taking measurements, means that many ground-based solar flare observations are marred with bad seeing.
To combat this, we propose a method for dedicated flare seeing correction based on training a deep neural network to learn to correct artificial seeing from flare observations taken during good seeing conditions.
This model uses transfer learning, a novel technique in solar physics, to help learn these corrections.
Transfer learning is when another network already trained on similar data is used to influence the learning of the new network.
Once trained, the model has been applied to two flare datasets: one from AR12157 on 2014/09/06 and one from AR12673 on 2017/09/06.
The results show good corrections to images with bad seeing with a relative error assigned to the estimate based on the performance of the model.
Further discussion takes place of improvements to the robustness of the error on these estimates.
\end{abstract}


%%%%%%%%%%%%%%%%%%%%%%%%%%%%%%%%%%%%%%%%%%%%%%%%%%

%%%%%%%%%%%%%%%%% BODY OF PAPER %%%%%%%%%%%%%%%%%%

\section{Introduction}
\label{sec:intro}
Atmospheric scintillation is ubiquitous in ground-based astronomy.
This poses a problem for all observers, particularly those studying highly variable phenomena.
State-of-the-art observing facilities utilise adaptive optics (AO) systems in their optical path to correct for the wavefront deviations introduced by the atmosphere.
However when the seeing conditions are particularly bad; the object being observed evolves faster than the speed of the AO system; or, the field-of-view is much larger than the isoplanatic patch, post-processing techniques must be introduced to correct for seeing.
The two most common post-processing techniques used in solar physics are Speckle interferometry and Phase Diversity (PD) methods.

Speckle interferometry is the process which involves dividing the field of view into subfields that are smaller than the turbulence coherence length and correcting each individually.
As a result, this depends on accurate statistics of the atmospheric turbulence and, consequently, requires 100s of frames to estimate a diffraction-limited reconstruction \citep{von_der_luhe_solar_1987,von_der_luhe_speckle_1993}.
For the most dynamic of processes (e.g. solar flares), the evolution time will be shorter than the cumulative length of the exposures required to obtain the frames necessary for the reconstruction. Therefore the atmospheric parameters will need to be estimated from a number of consecutive frames which is much less than optimal for the algorithm. 
This will lead to greater uncertainty in the atmospheric parameters and a poorer restoration, as a result.
% For the most dynamic of processes (e.g. solar flares), the camera read-out speed becomes the limiting factor in a Speckle reconstruction.
% \jaa{Therefore, for flares (where the evolution time is likely to be less than the exposure), the scene is unlikely to be static over the exposure leading to a lower number of usable frames and a worse estimate of the atmospheric statistics.}
% The number of frames obtained within the exposure time will be smaller for more dynamical processes leading to a worse restoration by the Speckle method due to increased uncertainty in the atmospheric statistics. \lf{[This is a bit unclear to me - the `exposure time' is how long the shutter is open for, so are you thinking of multiple readouts within one exposure (e.g. with a CMOS or rolling-shutter CCD). Is it clearer to say that a single speckle image requires a certain number of frames, and for very dynamical processes the scene can change a lot over that set of frame?]}

PD methods jointly estimate the restored image and the distortions responsible for the aberrated image in a maximum likelihood estimation.
The state-of-the-art PD method in solar physics is Multi-Object Multi-Frame Blind Deconvolution \citep[MOMFBD; ][]{van_noort_solar_2005}.
MOMFBD implements a simple model of the optics and detectors used in the observations, eliminating the need to rely on the atmospheric statistics as in Speckle reconstruction.
As explained in \citet{van_noort_solar_2005}, PD methods work best when constrast is high, noise is low, and exposure time is short.
This is difficult to achieve in narrowband solar observations and, as a result, wideband data collected simultaneously must be used to aid in the MOMFBD restoration.
It is this that poses the biggest problem for the restoration of flare data.
Chromospheric energy deposition in a flare is mostly seen through the enhancement of optical and near-infrared (NIR) spectral lines and not necessarily strong continuum enhancements \citep{fletcher_observational_2011}.
This can lead to the objects being studied looking very different in the wideband and narrowband observations.
Given that the wideband is used to help the optimisation of the restoration, in cases where there is no continuum enhancement in a flare, it can actually be a hindrance to the restoration.

Furthermore, both Speckle and PD methods have a limit to their restoration capabilities (as all methods will).
This is detrimental to flare observations due to their sporadic nature meaning observers cannot wait for optimal seeing conditions to observe.
For these reasons, we propose a dedicated flare seeing-correction tool based on training a deep neural network on diffraction-limited narrowband flare data synthesised with artificial seeing.

Deep learning is the science of using deep neural networks to learn previously not-easily-programmable tasks; to learn computationally-expensive tasks to increase efficiency; and to gain insight into data via the use of data-driven models.
The use of deep learning in solar physics has already been applied to solar image restoration for increasing the speed at which MOMFBD restores images \citep{asensio_ramos_real-time_2018} and for the estimation of the point-spread function to do multi-frame blind deconvolution \citep{asensio_ramos_learning_2020}.
Furthermore, the correction of noisy narrowband polarisation data has been achieved using deep learning \citep{diaz_baso_solar_2019}.
Here we present a tool for correcting for seeing in observations of Stokes I in narrowband observations of solar flares, utilising a trained deep neural network to learn the mapping from seeing-plagued data to diffraction-limited data without the constraints or Speckle of PD methods.
% \citep[it is hypothesised that the seeing in Stokes Q, U, and V can be corrected for using the method in ][]{diaz_baso_solar_2019}. \lf{I moved this this statement in () but it seems a little out of place at this ppint in the paper. Maybe move to somewhere in the discussion?}

The structure of the paper is as follows: Section~\ref{sec:model} introduces the model used to generate synthetic seeing and a description of the data used to train the deep neural network; Section~\ref{sec:nna} highlights the architecture and training of the deep neural network with some validation data shown; Section~\ref{sec:2014} is the application of the model to observations from the dataset described in Section~\ref{sec:gtd} with natural bad seeing; Section~\ref{sec:2017} shows how well the model performs on data from the X9.3 flare SOL2017-09-06T12:03; and Section~\ref{sec:conc} is the conclusions.

\section{Seeing Model}
\label{sec:model}

The following outlines the synthetic seeing model that will be used to generate the training set for the neural network algorithm.
This synthetic seeing model will be applied to images which we assume are corrected completely by the adaptive optics and data reduction pipelines, 
%such that any synthetic seeing we add will not be 
and are not affected by residual seeing (i.e. the images are diffraction-limited).

Atmospheric seeing is the refraction of light as it travels through the Earth's atmosphere due to the turbulent nature of the atmosphere's refractive index.
This turbulence is caused by random variations in the density and temperature structure of the atmosphere.
%% LF sentence has been moved from here.
This greatly impacts the resolution of observations as images are  subject to the following: 

\begin{equation}
    \label{eq:im}
    O = I \ast P_{\text{atmos}} + G
\end{equation}

where $\ast$ denotes the convolution, $I$ is the diffraction-limited image, $P_{\text{atmos}}$ is the point-spread function (PSF) of the seeing, $G$ is random Gaussian noise \citep[for a discussion on the use of Gaussian noise see][]{van_noort_solar_2005}, and $O$ is the observed image.
The effect on astronomical imaging is that regardless of telescope aperture size, images appear as though observed through a telescope with effective aperture size equal to the Fried parameter, $r_{0}$ (Equation~\ref{eq:r0}).
The construction of the atmospheric PSF will provide a basis for applying synthetic seeing to diffraction-limited images for the network to learn from.
The general form for the PSF of the atmosphere is given by the Hankel transform of the Modulation Transfer Function (MTF)

\begin{equation}
    \label{eq:psf}
    P_{\text{atmos}} (\rho) = \int_{0}^{\infty} J_{0} (\rho\nu) \exp{\left \{-0.5 ~D_{S}(\nu) \right \}} \nu d\nu
\end{equation}

where $D_{S} (\rho)$ is the two point correlation (structure) function between the phase of two wavefronts in the telescope focal plane. $J_{0} (\rho\nu)$ is the zeroth order Bessel function and $\nu$ is spatial frequency.

To find a form for this structure function, the assumption is made that the Earth's atmosphere can be modelled as a medium with smoothly varying turbulence \citep{tatarski_wave_2016}.
Then, the structure function is written:

\begin{equation}
    \label{eq:sf}
    D_{S} (\rho) = 2.91 k^{2} \rho^{5/3} \int_{\vec{\ell}} C_{n}^{2} (\vec{r}) ~d\vec{r}
\end{equation}

where $k ~= ~2\pi/\lambda$ is the wavenumber of the light observed, $\rho$ is the Euclidean distance in the sky and $C_{n}^{2}$ is the profile describing the structure of the atmosphere at a point $\vec{r}$ (i.e. this encompasses the turbulent nature of the refractive index of the atmosphere).
$\vec{\ell}$ is the path taken by a photon through the atmosphere.
For a photon incident on the ``top'' of the Earth's atmosphere (the point where the medium becomes turbulent) at an angle $\theta$ to the normal of the atmosphere, the $C_{n}^{2}$ profile can be written:

\begin{equation}
    \label{eq:cn2}
    \int_{\vec{\ell}} C_{n}^{2} (\vec{r}) ~d\vec{r} = \sec ~\theta ~\int_{0}^{\ell} C_{n}^{2} (z) ~dz
\end{equation}{}

where the $z$-direction represents the direction of the normal to the atmosphere and the limit $z$=0 corresponds to the top of the atmosphere, and $z$=$\ell$ to the total distance travelled by the photon.
Equation \ref{eq:sf} can then be rewritten as

\begin{equation}
    \label{eq:sf2}
    D_{S} (\rho) = 2.91 k^{2} \rho^{5/3} \sec ~\theta ~ \int_{0}^{\ell} C_{n}^{2} (z) ~dz
\end{equation}

The Fried parameters is then written explicitly as \citep{fried_optical_1966}

\begin{equation}
    \label{eq:r0}
    r_{0} = \left ( 0.423 k^{2} \sec ~\theta ~\int_{0}^{\ell} C_{n}^{2} (z) ~dz \right )^{-3/5}
\end{equation}

This can then be used to simplify Equation \ref{eq:sf2} to

\begin{equation}
    \label{eq:sf3}
    D_{S} (\rho) = 6.88 \left ( \frac{\rho}{r_{0}} \right )^{5/3} = 6.88 \left ( \frac{\lambda \nu}{2\pi r_{0}} \right )^{5/3}
\end{equation}

where $\lambda$ here represents the air wavelength of the light observed and the spatial frequency $\nu$ is expressed in units of radians of phase per radian field of view \citep{racine_telescopic_1996}.

The form of the structure function given by Equation~\ref{eq:sf3} is used as it eliminates the need to choose a model for the $C_{n}^{2}$ profile and instead, the Fried parameter becomes a free parameter in the model with a variety of different values explored.

As the image is degraded in quality to equivalent to one taken with an aperture of diameter $r_{0}$, the angular size of the PSF in the sky can be found using: 

\begin{equation}
    \label{eq:angsize}
    \alpha = 2.021\times10^{5} ~\times ~ \frac{\lambda}{r_{0}}
\end{equation}

where $\alpha$ is measured in arcseconds.
The size of the PSF in detector pixels ($n_{\text{pix}}$) can then be calculated by dividing by the angular size of a single pixel ($\alpha_{\text{pix}}$)

\begin{equation}
    \label{eqn:npix}
    n_{\text{pix}} = \frac{\alpha}{\alpha_{\text{pix}}}.
\end{equation}

$n_{\text{pix}}$ is then the size of the PSF array to be convolved with the image (under the diffraction-limited assumption).
The PSF is then populated using Equations~\ref{eq:psf}~and~\ref{eq:sf3}.

\subsection{Generating Training Data}
\label{sec:gtd}
A range of Fried parameters  $r_{0}$=\{1, 2.5, 5, 7.5, 10, 12.5, 15\}cm is used to generate many different PSFs to convolve with the good seeing images following Equation~\ref{eq:im}.
This creates a diverse training dataset for the neural network to learn from.

The data used are taken with the Swedish 1m Solar Telescope's CRisp Imaging SpectroPolarimeter (SST/CRISP) instrument \citep{scharmer_1-meter_2003, scharmer_comments_2006, scharmer_crisp_2008}.
CRISP is a dual Fabry-P\'{e}rot interferometer capable of narrowband imaging spectropolarimetry and wideband imaging.
Imaging spectroscopy data is used in the training of the network with observations in two spectral lines: H$\alpha$ and Ca~\textsc{ii} $\lambda$8542.
The observations are of three flares: the M1.1 two-ribbon solar flare SOL20140906T17:09 which took place in NOAA AR 12157 with heliocentric coordinates (-732$^{\prime \prime}$, -302$^{\prime \prime}$); an X2.2 event SOL20170906T09:10 which took place in NOAA AR 12673 with heliocentric coordinates (537$^{\prime \prime}$, -222$^{\prime \prime}$); and an X9.3 event SOL201709-06T12:02 taking place in the same AR as SOL20170906T09:10.
For SOL2014-09-06T17:10, the H$\alpha$ data is sampled at 15 wavelength points in intervals of 200m\AA{} from the line core, and the Ca~\textsc{ii} data consist of 25 wavelength points sampled at 25 wavelengths in intervals of 100m\AA{} from the line core.
This data is made publicly available via the F-CHROMA solar flare database \citep{cauzzi_f-chromaflare_2014}\footnote{\url{https://star.pst.qub.ac.uk/wiki/doku.php/public/solarflares/start}}.
For SOL20170906T09:10 and SOLT20170906T12:02, the H$\alpha$ data is taken at 13 wavelength points with the line core more densely sampled than the wings. The Ca\textsc{ii} data is similarly sampled but for 11 wavelength points.
All data has been preprocessed using the CRISPRED data reduction pipeline \citep{de_la_cruz_rodriguez_crispred:_2015} which includes all alignment, instrument calibration, and image restoration using MOMFBD.
Therefore, the ground truth to be recovered makes the assumption that images without bad seeing conditions are completely corrected for seeing and other aberrations by the CRISPRED pipeline.
%-- in other words in this instance the network will learn to `remove' seeing to a level equivalent to what is achievable with AO seeing correction, but will do so across the entire field-of-view and not just the limited isoplanatic patch.

\begin{figure*}
    \centering
    \includegraphics[width=\textwidth]{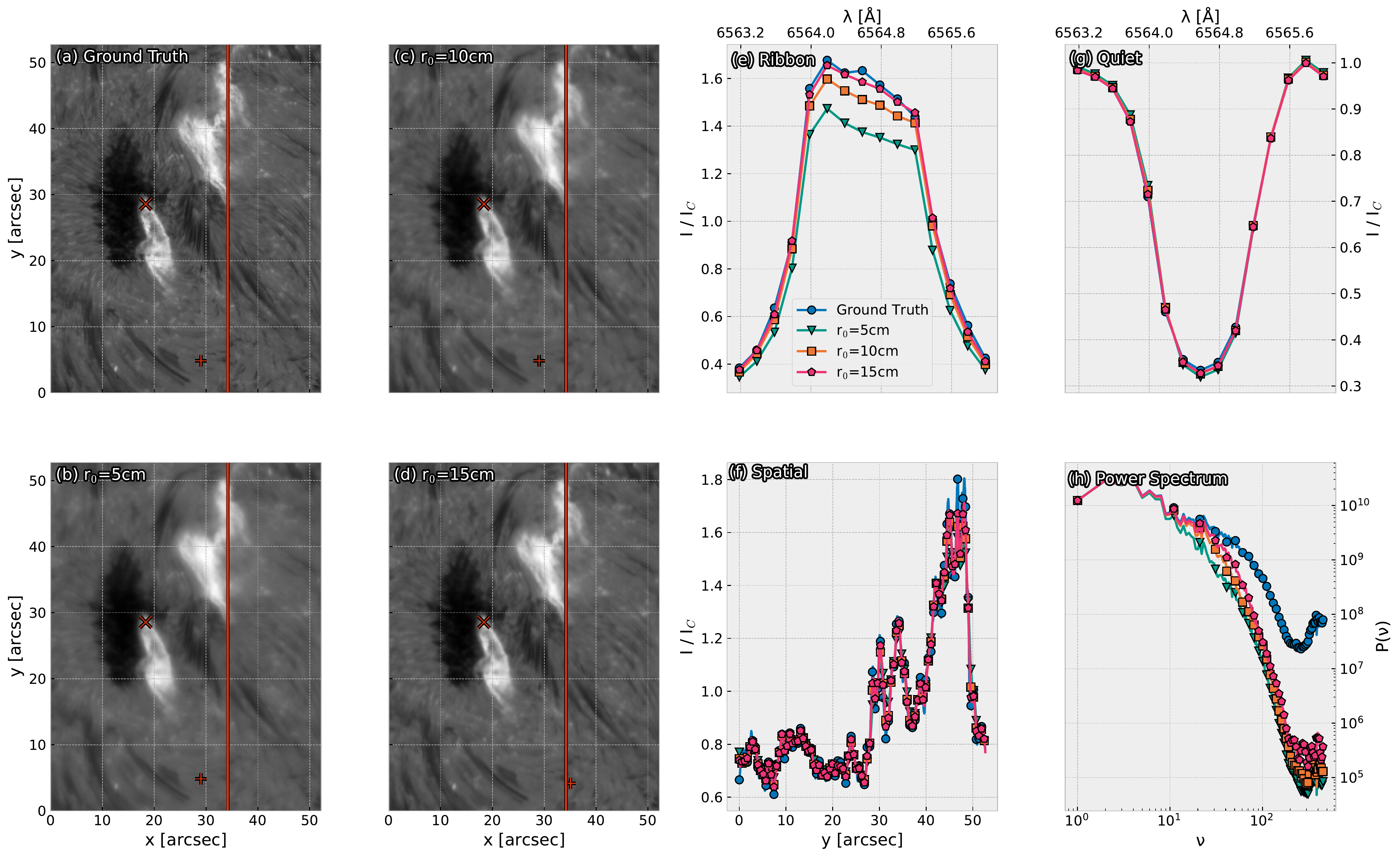}
    \caption{Example of the seeing model described in Section~\ref{sec:model} applied to one of the good seeing observations from the dataset described in Section~\ref{sec:gtd}, particularly, this observation is from 17:11:33UTC approximately 2 minutes after the flare soft X-ray peak, shown in panel (a).
    The image used here for demonstration is of the H$\alpha$ blue wing at $\Delta \lambda$=-600m\AA{}.
    The seeing model is applied for three different Fried parameters as can be seen in (b) $r_{0}$=5cm, (c) $r_{0}$=10cm, and (d) $r_{0}$=15cm.
    (e), (f), and (g) show the change in the spectral line on the flare ribbon, a spatial slice, and the spectral line in a quieter part of the atmosphere, respectively.
    The flare ribbon line is indicated by the cross in panels (a)--(d) with the quiet point being the plus sign and the slice shown by the vertical line.
    (h) shows the azimuthally-averaged power spectrum across the images.
    In panels (e), (f), (g), and (h), the circles correspond to the ground truth, the triangles to $r_{0}$=5cm, the squares to $r_{0}$=10cm, and the pentagons to $r_{0}$=15cm.}
    \label{fig:sm}
\end{figure*}

An example of the seeing model for three different values of the Fried parameter ($r_{0}$=5, 10, 15~cm) is shown in Figure~\ref{fig:sm}.
Figure~\ref{fig:sm}(a) shows the observation from 17:11:33UTC from the dataset described above (approximately 2 minutes after the flare soft X-ray peak) taken in the H$\alpha$ blue wing at $\Delta \lambda$=-600~m\AA{}.
Figures~\ref{fig:sm}(b)--(d) then show seeing corresponding to $r_{0}$=5~cm, 10~cm, 15~cm, respectively, applied to the observation.
Two points are then selected: one on the eastern flare ribbon indicated by the cross in Figures~\ref{fig:sm}(a)--(d) and one in the quiet atmosphere represented by the plus in the same figures. The spectra of these points in each of the four cases are plotted in Figure~\ref{fig:sm}(e) for the ribbon and Figure~\ref{fig:sm}(g) for the quiet atmosphere.
This indicates that worsening seeing will result in a reduction in the intensity of bright features due to the spatial smearing of the intensity.
Figure~\ref{fig:sm}(f) shows the intensity variation with $y$ for a fixed $x$ indicated by the vertical line in Figures~\ref{fig:sm}(a)--(d).
Figure~\ref{fig:sm}(f) solidifies the previous points, showing that peaks and troughs of intensity are lost as seeing worsens.
Also small-scale features are lost due to bad seeing, which is apparent as curves in Figure~\ref{fig:sm}(f) become smoother for worse seeing.
The power spectrum is also calculated for each image shown in Figure~\ref{fig:sm}(h), which further conveys the loss of small scale features as the power in the higher frequencies is substantially reduced as seeing worsens.

\section{Neural Network Approach}
\label{sec:nna}
The following section outlines the deep neural network (DNN) architecture trained and used to correct for atmospheric seeing, and a description of how the network is trained.

\subsection{Architecture}
\label{sec:nnaa}
\begin{figure*}
    \centering
    \includegraphics[width=0.99\textwidth]{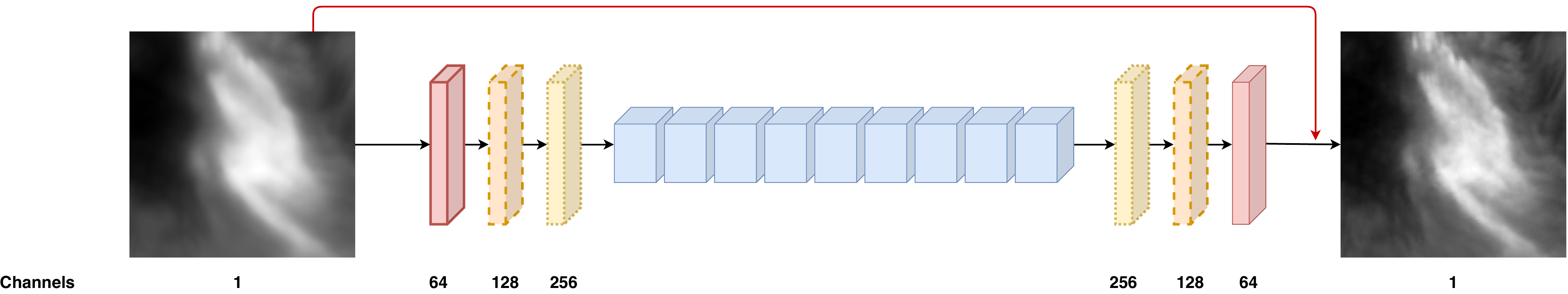}
    \caption{Schematic of the neural network used to learn the seeing correction.
    This network consists of six convolutional layers and nine residual layers.
    The picture on the left of the network shows the input which is an image from the dataset generated in Section~\ref{sec:gtd} of an image imbued with synthetic seeing.
    The picture on the right is the ground truth the network is trying to recover.
    The first block (with the solid lines) is a convolutional layer using a 7$\times$7 kernel and generating 64 feature maps.
    The block with the dashed lines downsamples the feature maps produced by the first block by a factor of 2 using a strided convolution of 3$\times$3 kernel and produces 128 feature maps.
    The dotted line block downsamples the feature maps by a further factor of 2 using a strided convolution of 3$\times$3 kernel and produces 256 feature maps.
    The shorter blocks in the middle are the residual layers which all consist of 3$\times$3 kernel convolutions with 256 feature maps.
    The inner structure of the residual layers is shown in Figure~\ref{fig:rb}.
    The next dotted line block upsamples the feature maps by a factor of 2 using nearest-neighbour interpolation and reduces the number of feature maps to 128.
    The second dashed line block then upsamples by a further factor of 2 using the same method while reducing the number of feature maps to 64.
    The last block in the network is a convolutional block which reduces the number of feature maps to the number of output channels using a 7$\times$7 kernel convolution before passing the output through a hyperbolic tangent (tanh) function.
    This is then combined with the input to the network (red arrow) to produce the output of the network.
    In each of the convolutional and residual layers the noramlisation is batch normalisation and the activation is ReLU.}
    \label{fig:shauna}
\end{figure*}
\begin{figure*}
    \centering
    \includegraphics[width=0.9\textwidth]{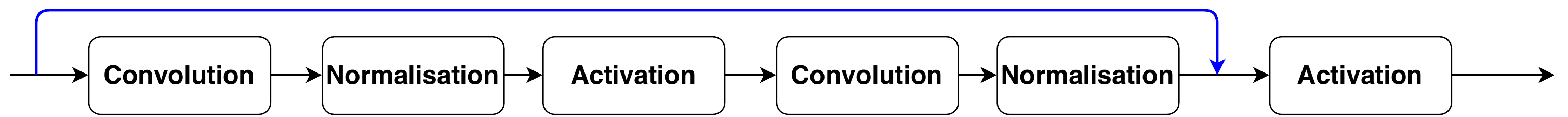}
    \caption{Inside of a residual block. This consists of two convolution layers applied to the input like traditional convolutional neural networks but with a skip connection (blue arrow) adding the input of the layer to the output before the second activation.
    This allows residual networks to be deeper than traditional networks as it prolongs the onset of the vanishing gradient problem.}
    \label{fig:rb}
\end{figure*}

The DNN architecture used is illustrated in Figure~\ref{fig:shauna} and inspired by the generator network used in \citet{kupyn_deblurgan:_2017}.
The network follows an encoder-decoder framework wherein the input data -- in this case, the image with bad seeing -- is downsampled to a lower-dimensional, abstract representation of itself that can be reconstructed without the bad seeing by the learned network by upsampling the representation at the other end of the network.
This is accomplished using a combination of convolutional layers (consisting of convolution, normalisation and activation) and residual layers \citep[Figure~\ref{fig:rb};~][]{he_deep_2015}.
The normalisation used in the convolutional and residual layers is batch normalisation \citep{ioffe_batch_2015}, and the activation used is rectified linear unit \citep[ReLU;~][]{nair_rectified_2010}.
The convolutional layers are used for the downsampling and upsampling of the data while the residual layers learn the complexities of the abstract representation.
Nine residual layers was found to be the optimal number to learn to correct for the bad seeing, with three convolutional layers either side of the residual layers to perform the down/upsampling.

The first convolutional layer (shown with emboldened vertices in Figure~\ref{fig:shauna}) convolves the image with a 7$\times$7 kernel and transforms the input to 64 feature maps.
The dashed line layer convolves these feature maps with a 3$\times$3 kernel, downsampling their dimension by a factor of 2 and doubling the number of feature maps to 128.
The dotted line layer convolves the 128 feature maps in the same way as the previous layer, downsampling the feature maps by a factor of 2 and doubling the number of feature maps to 256.

After this, these feature maps are passed to the nine residual layers, shown as the shorter blocks in Figure~\ref{fig:shauna}.
Each of these layers have the structure shown in Figure~\ref{fig:rb}.
The convolution kernel sizes are all 3$\times$3 with each residual layer keeping the number of feature maps at 256.

Subsequently, the feature maps are given to the second dotted line layer which upsamples the feature maps by a factor of 2 using nearest-neighbour interpolation and reduces the number of feature maps by a factor of 2 to 128 using a convolution with 3$\times$3 kernel.
Then, the second dashed line layer follows the same process resulting in there being 64 feature maps a factor of 2 larger than before being passed to the final layer.
The final layer transforms the feature maps to the number of output channels (in this case, 1) using a convolution with 7$\times$7 kernel.
The output of this layer is then operated on by a hyperbolic tangent (tanh) function before being combined with the input to the network (shown by the red arrow in Figure~\ref{fig:shauna}).
Being combined with the input is what \citet{kupyn_deblurgan:_2017} coined as a ``ResOut" connection.
The philosophy behind this is that the network is trying to learn some function $f$ which maps an input with bad seeing to an output with good seeing.
Given that $f$ depends on the input, there exists some residual function, $H$, such that
\begin{equation}
    \label{eq:res}
    H(x) = f(x) - x
\end{equation}
where the input to the network has been denoted by $x$.
Therefore, adding the input at the end of the network allows it to learn only the residual $H$ which may be easier to learn than the function $f$.\footnote{This is also the basis of how residual layers work and why networks can be much deeper using them.}

\subsection{Training}
\label{sec:nnat}

The network described in Section~\ref{sec:nnaa} is then trained using the data in Section~\ref{sec:gtd} with the images synthesised with seeing as the input to the network and the corrected images as the output.
% The data used are taken with the Swedish 1m Solar Telescope's CRisp Imaging SpectroPolarimeter (SST/CRISP) instrument \citep{scharmer_1-meter_2003, scharmer_comments_2006, scharmer_crisp_2008}.
% CRISP is a dual Fabry-P\'{e}rot interferometer capable of narrowband imaging spectropolarimetry and wideband imaging.
% Imaging spectroscopy data is used in the training of the network with observations in two spectral lines: H$\alpha$ and Ca~\textsc{ii} $\lambda$8542.
% The observations are of the M1.1 two-ribbon solar flare SOL20140906T17:09 which took place in NOAA AR 12157 with heliocentric coordinates (-732$^{\prime \prime}$, -302$^{\prime \prime}$).
% The H$\alpha$ data is sampled at 15 wavelength points in intervals of 200m\AA{} from the line core, and the Ca~\textsc{ii} data consist of 25 wavelength points sampled at 25 wavelengths in intervals of 100m\AA{} from the line core.
% The data is made publicly available via the F-CHROMA solar flare database \citep{2014AAS...22412339C}\footnote{\url{https://star.pst.qub.ac.uk/wiki/doku.php/public/solarflares/start}}, where it has been preprocessed using the CRISPRED data reduction pipeline \citep{de_la_cruz_rodriguez_crispred:_2015}.
% Therefore, the ground truth to be recovered makes the assumption that images without bad seeing conditions are completely corrected for seeing and other aberrations by the CRISPRED pipeline.

Aside from generation of a good training dataset, the key to training a network is to use the correct loss function to track how well the network is doing.
In this case, the loss function takes the form of two individual loss functions in a linear combination: perceptual loss and mean square error (MSE) loss

\begin{equation}
    \label{eq:totalL}
    \mathcal{L} = \mathcal{L}_{P} + \mathcal{L}_{\text{MSE}}
\end{equation}

Perceptual loss \citep[introduced by][]{johnson_perceptual_2016} is a measure of similarity between two images based on how they are perceived by a different neural network from the one being trained.
This is an example of transfer learning: the process of using a previously trained neural network to influence the learning of a new network.
The network from \citet{armstrong_fast_2019} (henceforth referred to as Slic) is used here due to it being trained to classify features in the solar atmosphere.
The argument is that a network trained sufficiently well on recognising features should produce the same feature maps for two identical images.
Therefore, using a measure of the difference of these features maps produced deep within the Slic network will give a measure of the similarity in the two images.
This works by taking the network generated image $I_{G}$ and the ground truth image $I_{S}$ and applying Slic to them.
The output is then cut after the eighth layer and the feature maps compared using a mean square error metric.
This can be written:

\begin{equation}
    \label{eq:percL}
    \mathcal{L}_{P} = \frac{1}{W_{j} H_{j}} \sum_{x=0}^{W} \sum_{y=0}^{H}\left(\phi_{j}\left(I_{S}\right)_{x, y}-\phi_{j}\left(I_{G}\right)_{x, y}\right)^{2}
\end{equation}

where $W_{j}$, $H_{j}$ are the width and height of the jth output layer of Slic, respectively.
$\phi_{j}$ is the function resulting from feeding the images through Slic and taking the output after the eighth layer.

The MSE loss is the N-dimensional Euclidean distance function squared where the data is N-dimensional.

\begin{equation}
    \label{eq:MSEL}
    \mathcal{L}_{\text{MSE}} = || I_{S} - I_{G} ||^{2}
\end{equation}

The MSE ensures that the magnitude of the reconstructions match similarly to the ground truth images.

The perceptual and MSE losses are then minimised simultaneously using the Adam optimiser \citep{kingma_adam:_2014} using minibatching and a variable learning rate following cosine annealing.
Minibatching consists of not using the entirety of the training and validation datasets while training the network.
Instead, 10\% of the training and validation data are used randomly per epoch for training.
This increases the speed of the epoch which can speed up the convergence of the network (diversity across the data will lead to better generation as the network does not see the same data every epoch and having more but quicker epochs leads to more parameter updates and thus faster learning).
In training, a batch size of 12 is used with 100 minibatches per epoch for the training data and 10 minibatches for the validation data.
Cosine annealing \citep{loshchilov_sgdr:_2016} is a method for dynamically changing the learning rate of the system every epoch following:
\begin{equation}
    \label{eq:ca}
    \eta_{t}=\eta_{\min }+\frac{1}{2}\left(\eta_{\max }-\eta_{\min }\right)\left(1+\cos \left(\frac{T_{\text{cur}}}{T_{\max }} \pi\right)\right)
\end{equation}
where $\eta_{t}$ is the current learning rate, $\eta_{\min}$ is the minimum learning rate, $\eta_{\max}$ is the starting learning rate, $T_{\text{cur}}$ is the current epoch number, and $T_{\max}$ is the number of epoch to get to the minimum learning rate.
This method allows the exploration of local minima by decreasing the learning rate from $\eta_{\max}$ to $\eta_{\min}$ over $T_{\max}$ epochs while allowing the network to escape from incorrect local minima as the learning rate is reset to $\eta_{\max}$ after $T_{\max}$ epochs.
The network here has $\eta_{\max}$ = 5$\times$10$^{-3}$, $\eta_{\min}$ = 1$\times$10$^{-6}$, and $T_{\max}$ = 100.
The network is trained on an NVIDIA Titan Xp for 1900 epochs.
The results are shown in Section~\ref{sec:res} below.

\subsection{Training Results}
\label{sec:res}
\begin{figure*}
    \centering
    \includegraphics[width=0.9\textwidth]{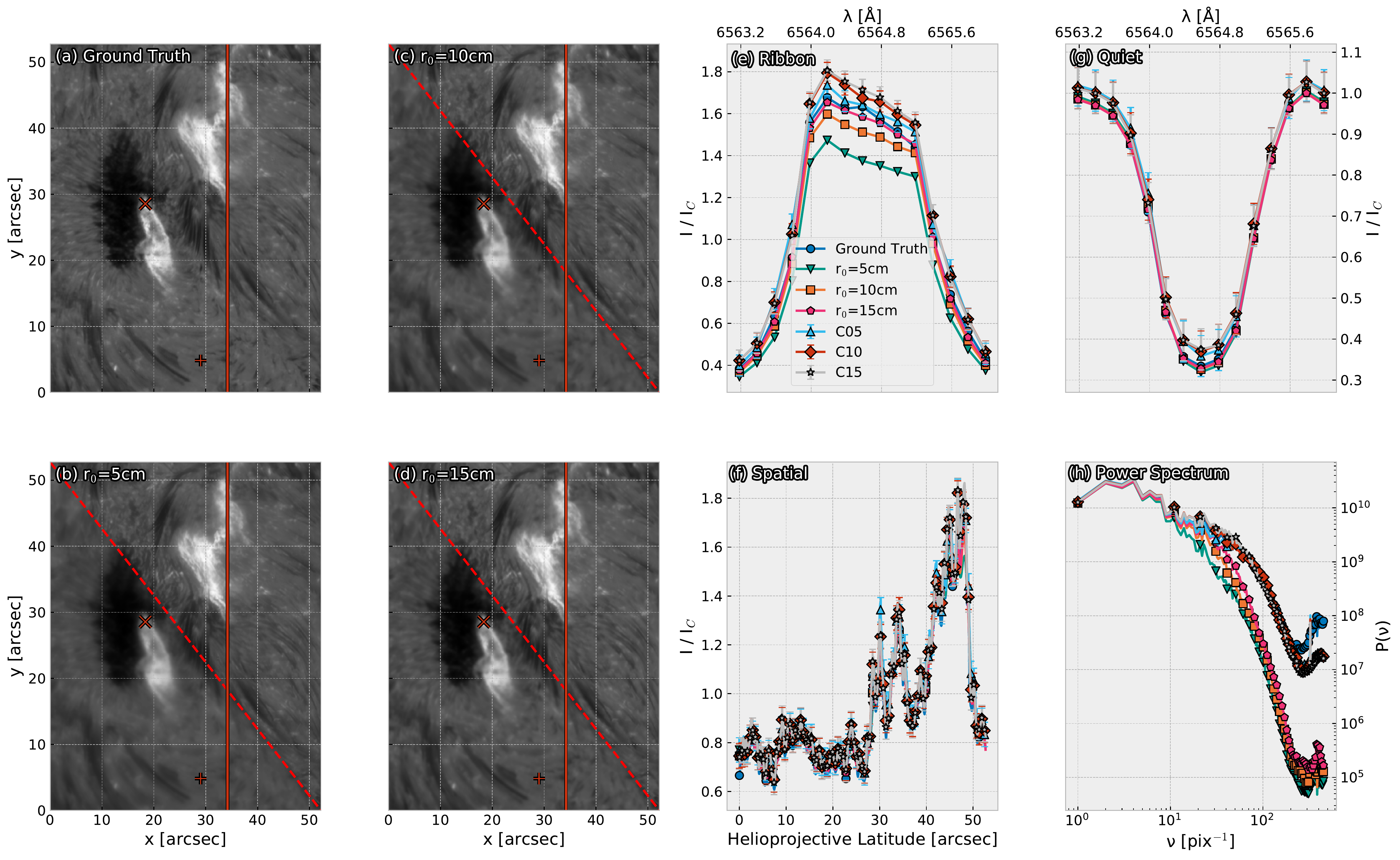}
    \caption{The results for applying the trained model to the data presented in Figure~\ref{fig:sm}.
    The layout of this figure is equivalent to the layout of Figure~\ref{fig:sm} with a few differences.
    Panels (b)--(d) show both the ground truth data contaminated with the artificial seeing (data below the dashed line) and the reconstructed data from the network (data above the dashed line).
    This demonstrates the reconstructive power of the trained network for images with artificial seeing as the small-scale features that can be reconstructed are visible by eye.
    Note that the reconstruction is worse for worsening seeing conditions which is to be expected from any algorithm of this kind.
    Panels (e)--(h) show the same as in Figure~\ref{fig:sm} with the profiles for each of these cases for the reconstructions also plotted and indicated in the legend as e.g. ``C05'' refers to correction of the data with bad seeing characterised by r$_{0}$=5cm, etc.
    The reconstructed profiles are also plotted with their error bars, showing that the reconstruction is at least within the error bar of the ground truth.}
    \label{fig:results}
\end{figure*}

To test the trained model, the data generated and shown in Figure~\ref{fig:sm} is evaluated by the trained model with a spectral, spatial and power spectrum comparison as in Figure~\ref{fig:sm}.
Given that the neural network model is an approximate fitting \citep[in line with the Universal Function Approximation Theorem;][]{cybenko_approximation_1989,lu_expressive_2017}, a formulation of an error on the estimate by the network is important.
As such, an \emph{ad hoc} error is calculated by evaluating the whole training set by the trained model and taking the average error obtained from Equation~\ref{eq:totalL}.
Other methods of error estimation for neural networks are discussed in Section~\ref{sec:conc} but not explored in this work.
The results of the reconstruction are shown in Figure~\ref{fig:results}.

Figure~\ref{fig:results}(b)--(d) show both the ground truth data (Figure~\ref{fig:results}(a)) contaminated with artificial seeing (the data below the dashed line) and the degraded data that has been reconstructed by the trained neural network model (the data above the dashed line).
This shows the reconstructive power of the trained model, with small scale structure recovery visible by eye.
As with all algorithms of this kind, the reconstructions are of better quality for better seeing conditions.

For more quantitative measures of the reconstruction, three profiles are compared: a spectral line profile on the flare ribbon; a spectral line profile in a quieter part of the atmosphere; and a spatial line profile which is a slice of constant $y$ shown by the cross, plus, and vertical line in Figure~\ref{fig:results}(a)--(d), respectively.
The on-ribbon spectral line profile is plotted in Figure~\ref{fig:results}(e).
The ground truth is indicated by the circular markers with the degraded data indicated by downward facing triangles for r$_{0}$=5cm, squares for r$_{0}$=10cm, and pentagons for r$_{0}$=15cm.
The reconstructed profiles are also plotted with their error bars and are indicated by upward facing triangles for the reconstruction of r$_{0}$=5cm, diamonds for r$_{0}$=10cm, and stars for r$_{0}$=15cm.
Despite the large error bars, each case is reconstructed well by the model with the ground truth falling within the error bars and a noticeable return of the unusual shape near the peak of the line.
Figure~\ref{fig:results}(g) show the results for the spectral line from the quieter part of the atmosphere.
This follows the same convention as Figure~\ref{fig:results}(e).
In this case, the reconstruction is somewhat worse regardless of the seeing conditions as the wings of the line still have a discrepancy compared with the ground truth intensity values.
This may be due to the focus of the trained model being subverted by the bright features with the lower contrast features not being as crucial in the reconstruction.
Figure~\ref{fig:results}(f) shows the slice of constant $y$.
This illustrates on the whole that the brighter and darker features are reconstructed well by the model as there is not much discrepancy between reconstruction and ground truth along the slice.

Figure~\ref{fig:results}(h) shows the power spectrum for the ground truth, each of the degraded images and the reconstructed images following the same descriptions as Figure~\ref{fig:results}(e)--(g).
The reconstructions show that the large to medium scale structure (up to $\nu$=10 pix$^{-1}$) within the field-of-view is almost perfectly reconstructed regardless of seeing conditions.
The rest of the spectrum for each reconstruction shows a tendency to reconstruct smaller features but not with the power they are represented by in the ground truth image.
This is noticeable towards the highest frequencies where length scales approach a single pixel, however, when the features are on scales of 10s of pixels their power is still restored well for all seeing conditions.
For example, when r$_{0}$=15cm, there is still a good reconstruction up to approximately $\nu$=45~pix$^{-1}$.
The shapes of the reconstructed power spectra are correct compared with the ground truth which suggests learning for a better convergence of the L2 loss may result in the restoration of the lost power.

This has demonstrated the flexibility and accuracy of the restoration when applied to diffraction-limited images contaminated with artificial seeing.
Next, the soundness of the model applied to images with no ground truth will be explored.
This is done for two separate flares: Section~\ref{sec:2014} explores the M1.1 flare that the network is trained on but uses images that naturally could not be corrected by the CRISPRED data pipeline fully and Section~\ref{sec:2017} performs reconstructions on data from an X9.3 solar flare SOL2017-09-06T11:53.

\section{Correcting Bad Seeing on the Most Explosive Active Region of Solar Cycle 24}
\label{sec:2017}
\begin{figure*}
    \centering
    \includegraphics[width=0.99\textwidth]{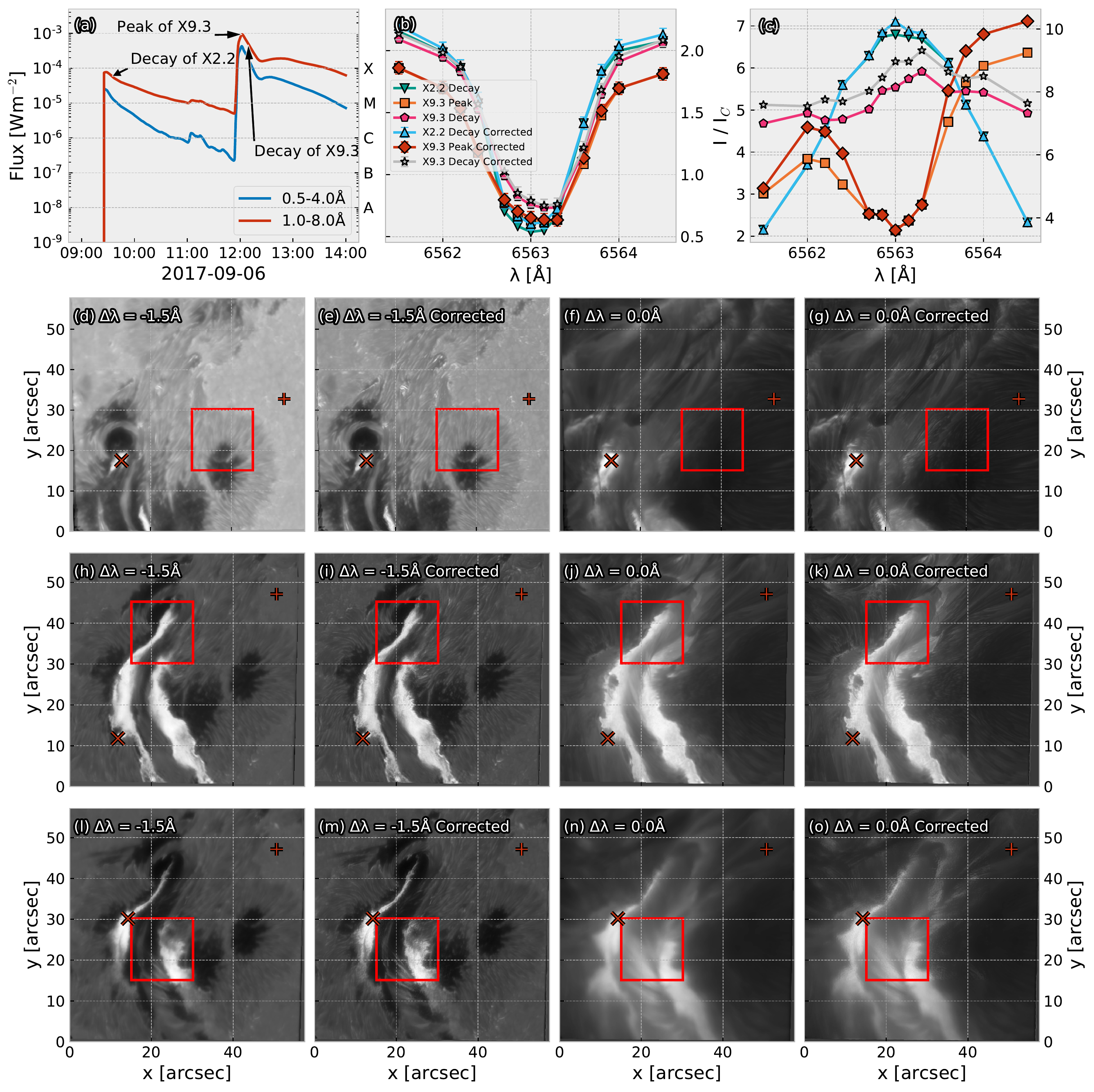}
    \caption{Panel (a) shows the GOES soft X-ray curve for the AR 12673 data, annotated to show where each observation corresponds to.
    (b) and (c) show spectra off and on the flare ribbon from the raw frames and the corrected spectra for each of the cases.
    Trained model applied to observations from AR 12673 for the decay phase of the X2.2 flare SOL20170906T09:10 (d)--(g); the soft X-ray peak of the X9.3 flare SOL20170906T12:02 (h)--(k); and the decay phase of SOL20170906T12:02 (l)--(o).
    In each row, the first panel is the observation before correction taken in the far blue wing of H$\alpha$ ($\Delta \lambda$=--1.5\AA{}); the second panel is the correction to the blue wing image; the third panel is the image in the line core before correction; and the last panel is the correction to the line core.
    The spectra shown are indicated in (d)--(o) using ``+'' and ``x'' for (b) \& (c), respectively.
    The boxes in panels (d)--(o) represent the subfields shown in Figure~\ref{fig:2017_out_zoom}.}
    \label{fig:2017_out}
\end{figure*}

The most energetic flare of Solar Cycle 24, SOL20170906T12:02, occurred in NOAA active region AR12673 with GOES class X9.3 (Figure~\ref{fig:2017_out}(a)).
This flare was observed by CRISP as described in Section~\ref{sec:gtd}.
Unfortunately, seeing conditions were so poor that most of the data is affected by residual seeing.
The true test for the neural network model is to see if it can reconstruct this data with accurate photometry across the field-of-view; perceptual similarities to their high resolution counterparts; and sensible spectral line reconstruction.

In Figure~\ref{fig:2017_out}, we show three examples of corrections made to H$\alpha$ observations with the neural network.
Figure~\ref{fig:2017_out}(a) shows the GOES soft X-ray lightcurves indicating the \emph{de facto} flare classification.
This is annotated to show the three different times the examples are from: in the decay phase of SOL20170906T09:10 at 09:34:26UTC; at the peak of SOL20170906T12:02 at 12:02:26UTC; and in the decay phase of SOL20170906T12:02 at 12:09:11UTC.
Figure~\ref{fig:2017_out}(d)--(g) show the SOL20170906T09:10 decay phase observation in the H$\alpha$ blue wing $\Delta \lambda$=--1.5\AA{}; the corrected observation in the blue wing; the observation in the H$\alpha$ line core; and the corrected observation in the line core, respectively.
Similarly, Figure~\ref{fig:2017_out}(h)--(k) show the peak of SOL20170906T12:02 and Figure~\ref{fig:2017_out}(l)--(o) show the decay phase of SOL20170906T12:02.

Each of Figure~\ref{fig:2017_out}(d)--(o) are annotated with a ``+'' and an ``x''.
The ``+'' indicates a point in a quieter part of the atmosphere with ``x'' indicating a point on the flare ribbons.
Correspondingly, the spectra from these points are shown in Figure~\ref{fig:2017_out}(b)\&(c).
In these panels, the downward facing triangles represent the spectral line before correction for the decay of SOL20170906T09:10 with the upward facing triangles representing the spectrum following correction; the square points show the line profile before correction for the peak of SOL20170906T12:02 with the diamonds showing the line profile post-correction; and the pentagons correspond to the profile before correction for the decay phase of SOL20170906T12:02 with the stars showing the profile post-correction.
The line profiles in Figure~\ref{fig:2017_out}(b) retain their shape when corrected with the intensity values in the wings (and, to a lesser extent, the core) increasing which we would expect as seeing will effectively ``smear'' light over many pixels causing a reduction in intensity in one pixel.
This correction also preserves asymmetry in the line profile and Doppler shifts which can be seen clearly due to the differences in wing intensities between the blue and red wing and the intensity-averaged line core not being equal to the emitted wavelength, respectively.
The line profiles in Figure~\ref{fig:2017_out}(c) show three very different line profiles on the flare ribbon depending on the time at which it is observed.
For the decay phase of SOL20170906T09:10, the line profile has small changes in the wings after correction but a larger change towards the line core.
The peak of SOL20170906T12:02 spectral line before correction appears as the characteristic twin-peaked H$\alpha$ profile (with a very broad red wing) with the correction implying that the blue wing should be stronger than in the raw observations.
The decay phase of SOL20170906T12:02 spectral line before and after correction maintains a similar shape with the intensities of the corrected profile being larger at every wavelength point.
The increases in intensity of each of these line profiles is to be expected by the same spatial ``smearing'' effect mentioned earlier but we would not expect the intensity to increase in every pixel (otherwise we would be introducing phantom photons to our observations).
To examine this further, another metric which we look to quantify the reconstruction is how well photon counts are preserved across both uncorrected and corrected images.
Since the data is level 2, i.e. corrected for the effects of instrumental and detector noise as well as alignment between channels and other data in the dataset and application of a seeing correction method, the unscaled intensity will have units of data numbers [DNs].
This means that to check conservation of photons, a sum over the field-of-view will suffice. %\lf{[in this dataset I guess that the signal is very much bigger than the CCD background (dark current etc) so this check is mostly for DNs related to photons. What if we were dealing with high background data? Just something to think about for the future]}
The results of this are shown in Table~\ref{tab:2017}.
In each frame, regardless of the event, the total DN decreases by a few percent.
This could mean that our neural network is under-compensating when correcting the flare ribbons or is over-compensating when correcting absorption features.
This is discussed more in Section~\ref{sec:conc}.

\begin{figure*}
    \centering
    \includegraphics[width=0.99\textwidth]{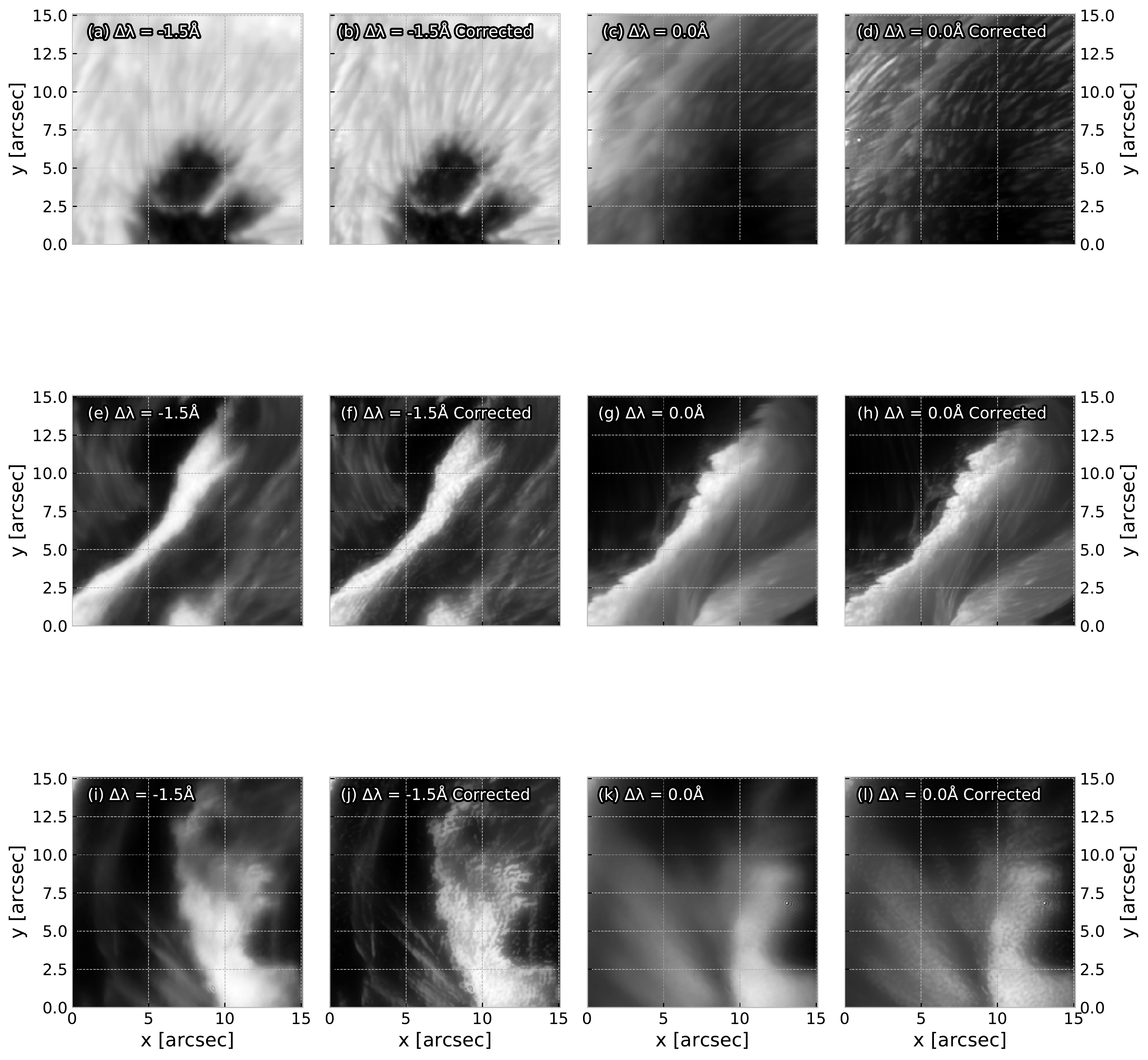}
    \caption{The subfields indicated by the boxes in Figure~\ref{fig:2017_out}.
    This shows the correction on the small scale features in the images for the three different observations of AR 12673 indicated in Figure~\ref{fig:2017_out}(a).
    Panels (a)--(d) show the model applied to a sunspot umbra/penumbra region in the decay phase of SOL20170906T09:10 in both the H$\alpha$ blue wing -- (a) \& (b) -- and H$\alpha$ line core -- (c) \& (d).
    Panels (e)--(h) show the application to the eastern flare ribbon at the peak of the SOL20170906T12:02 event following the same convention as the previous row.
    Similarly, panels (i)--(l) show the application to the western flare ribbon during the decay phase of SOL2017:0906T12:02.}
    \label{fig:2017_out_zoom}
\end{figure*}

The boxes in Figure~\ref{fig:2017_out}(d)--(o) reference the subfields shown in Figure~\ref{fig:2017_out_zoom}.
This is to illustrate how well our model recovers small-scale features in the flare ribbons and quieter parts of the Sun.
Figure~\ref{fig:2017_out_zoom}(a)--(d) shows part of the umbra/penumbra of AR 12673 for the decay phase of SOL20170906T09:10 both in the far blue wing -- panels (a) \& (b) -- and line core -- panels (c) \& (d) -- of H$\alpha$.
Figure~\ref{fig:2017_out_zoom}(e)--(h) shows the eastern flare ribbon in the prior format for the peak of SOL20170906T12:02 and Figure~\ref{fig:2017_out_zoom}(i)--(l) shows the western flare ribbon in the same format for the decay phase of SOL20170906T12:02.
This figure is for illustrative purposes and shows the good recovery of small-scale features even when the seeing is particularly bad, as is most prominently seen in the observation of the decay phase of SOL20170906T12:02.
% \lf{It looks a bit like additional small scales are generated in panel j; I think we should mention this and discuss later.}

\begin{table}
\centering
\caption{The percentage changes in the total number of DNs between the non-corrected and corrected images.
Negative percentage implies a lower number of DNs in the reconstruction and vice versa.}
\label{tab:2017}
\begin{tabular}{cccc}
\toprule
 $\Delta \lambda$ [\AA{}] &  X2.2 Decay[\%] &  X9.3 Peak [\%] &  X9.3 Decay [\%] \\
\midrule
                    -1.50 &           -2.50 &           -3.77 &            -3.62 \\
                    -1.00 &           -2.77 &           -3.57 &            -3.72 \\
                    -0.80 &           -3.12 &           -3.55 &            -3.42 \\
                    -0.60 &           -3.53 &           -3.10 &            -3.29 \\
                    -0.30 &           -4.60 &           -3.83 &            -3.38 \\
                    -0.15 &           -4.68 &           -3.75 &            -3.82 \\
                     0.00 &           -5.15 &           -3.88 &            -4.02 \\
                     0.15 &           -5.22 &           -3.71 &            -3.65 \\
                     0.30 &           -5.24 &           -3.39 &            -3.44 \\
                     0.60 &           -3.96 &           -3.31 &            -3.36 \\
                     0.80 &           -3.73 &           -3.43 &            -3.67 \\
                     1.00 &           -3.18 &           -3.52 &            -3.39 \\
                     1.50 &           -4.00 &           -3.80 &            -3.41 \\
\bottomrule
\end{tabular}
\end{table}

\section{Correcting Bad Seeing on the M1.1 Flare}
\label{sec:2014}
As with the AR12673 data, the AR12157 data described in Section~\ref{sec:gtd} contains both well-corrected data and data with seeing still present after the CRISPRED reconstruction.
% This is what inspired the training of a neural network to correct for seeing in these situations as a coherent time series of flare ribbon evolution will give insight into the nature of the energy deposition into the lower atmosphere.
The dataset can be used for both training and testing of the neural network model which provides some semblance of what the ground truth should be even when there is not a ground truth.

\begin{figure*}
    \centering
    \includegraphics[width=0.99\textwidth]{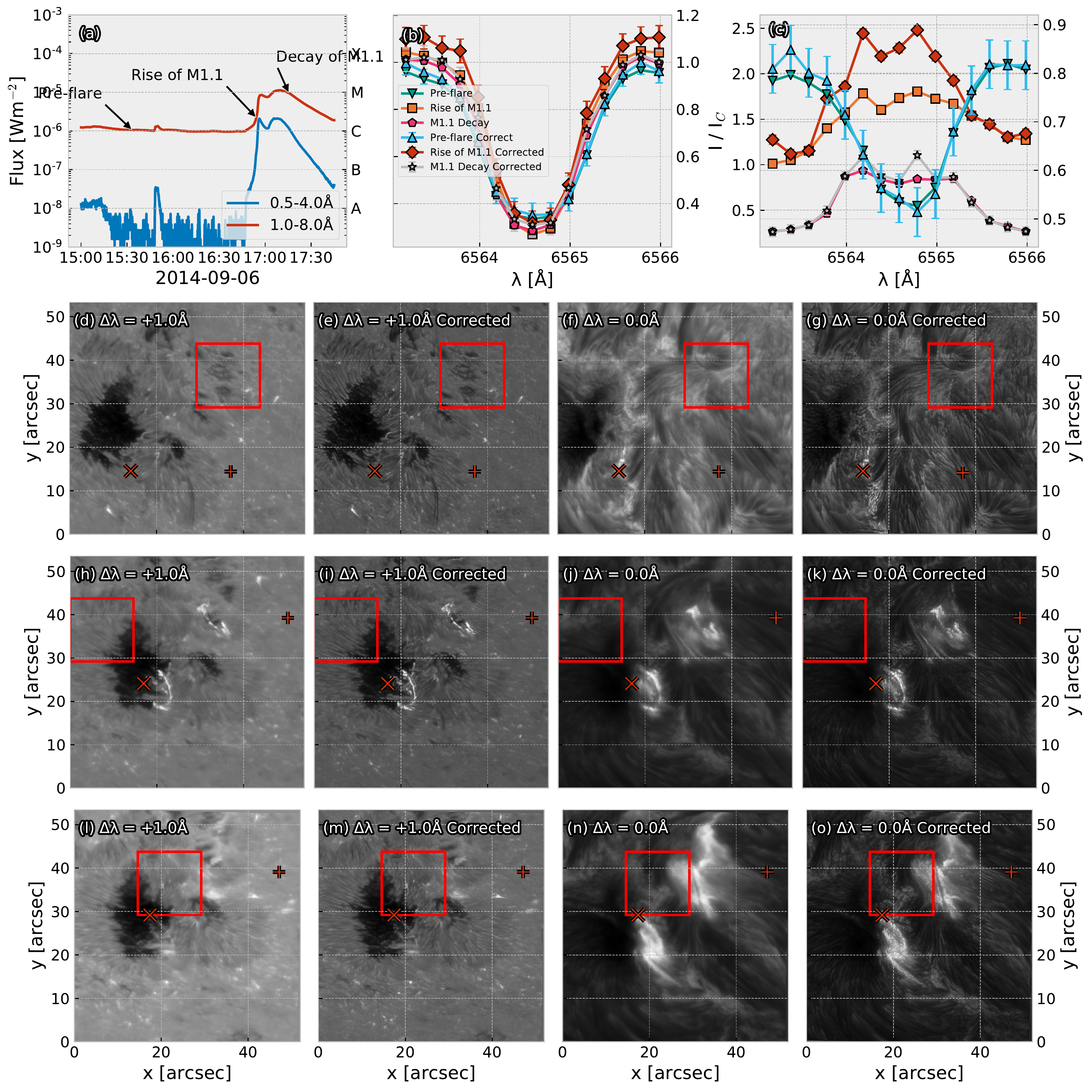}
    \caption{Panel (a) shows the GOES soft X-ray curve for the AR12157 data, annotated to show \lf{the time of each observation}.
    (b) and (c) show absorption and emission spectra, respectively from the raw frames, and the corrected versions for each of the cases.
    Trained model applied to observations from AR12157 for the pre-flare of SOL20140906T17:10 (d)--(g); the rise of the soft X-ray peak of SOL20140906T17:10 (h)--(k); and the decay of SOL20140906T17:10 (l)--(o).
    In each row, the first panel is the observation before correction, taken in the red wing of H$\alpha$ ($\Delta \lambda$=+1.0\AA{}); the second is the correction to the red wing image; the third panel is the image in the line core before correction; and the last panel is the corrected line-core image.
    The spectra shown are indicated in (d)--(o) using the ``+'' and ``x'' for (b) \& (c), respectively.
    The boxes in panels (d)-(o) represent the subfields shown in Figure~\ref{fig:2014_zoom}.}
    \label{fig:2014}
\end{figure*}

\begin{figure*}
    \centering
    \includegraphics[width=0.99\textwidth]{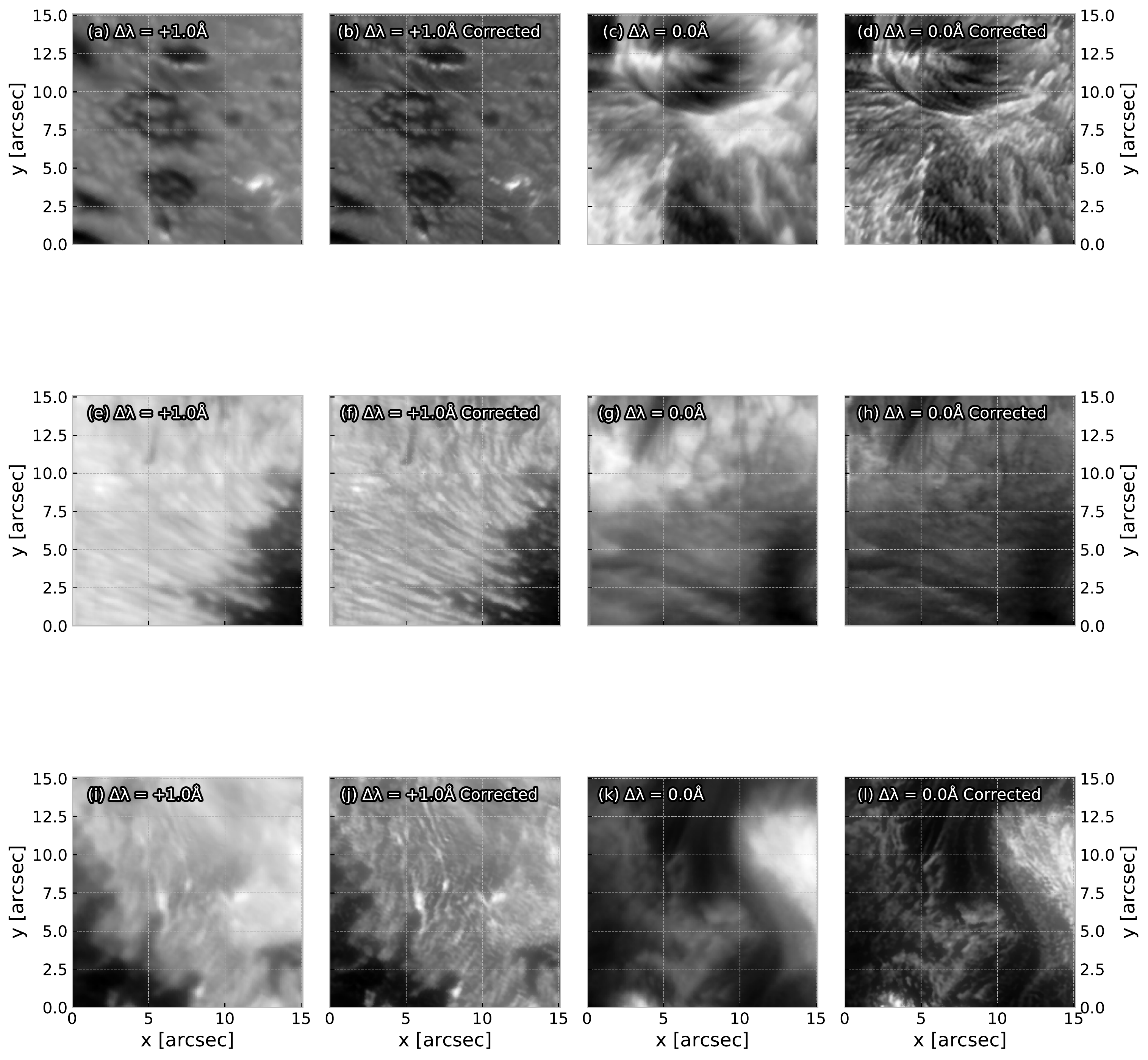}
    \caption{The subfields indicated by the boxes in Figure~\ref{fig:2014}.
    These show the correction on the small scale features in the image for the three different observations of AR12157 indicated in Figure~\ref{fig:2014}(a).
    Panels (a)--(d) show the model applied to part of AR12157 northwest of the main sunspot during the pre-flare of SOL20140906T17:10 in both the H$\alpha$ red wing -- (a) \& (b) -- and H$\alpha$ line core -- (c) \& (d).
    Panels (e)--(h) show the application to part of the sunspot umbra/penumbra during the rise of SOL20140906T17:10 following the same layout as the previous row.
    Likewise, panels (i)--(l) show the application to a region containing some sunspot penumbra and some of the northern flare ribbon during the decay of SOL20140906T17:10.}
    \label{fig:2014_zoom}
\end{figure*}

There are three examples in this section where the data contains bad seeing: one from the pre-flare phase, one from the rise of the soft X-ray peak, and one in the decay phase.
The results of the neural network training and testing are shown in Figures~\ref{fig:2014}~\&~\ref{fig:2014_zoom}.

Figure~\ref{fig:2014}(a) shows the GOES soft X-ray lightcurves indicating the flare classification and is annotated to show the three different times. The examples are: the pre-flare of SOL20140906T17:10 at 15:33:14UTC; during the rise of the soft X-ray peak of SOL20140906T17:10 at 16:54:13UTC; and in the decay of SOL20140906T17:10 at 17:15:24UTC.
Figure~\ref{fig:2014}(d)--(g) shows the pre-flare observation in the H$\alpha$ red wing $\Delta \lambda$=+1.0\AA{}; the corrected red wing observation; the H$\alpha$ line core observation; and the corrected line core observation, respectively.
Figure~\ref{fig:2014}(h)--(k) and Figure~\ref{fig:2014}(l)--(o) follow the same layout for the rise of the soft X-ray peak of SOL20140906T17:10 and the decay of SOL20140906T17:10, accordingly.

Each of Figure~\ref{fig:2014}(d)--(o) is annotated with a ``+'' and a ``x''.
The ``+'' indicates the spectra shown in Figure~\ref{fig:2014}(b) and the ``x'' the spectra shown in Figure~\ref{fig:2014}(c).
In these panels, the downward-facing triangles represent the spectral line before correction for the pre-flare observation and the upward facing triangles represent the post-correction spectrum; the square points show the line profile before correction for the rise of the soft X-ray peak of SOL20140906T17:10 with the diamonds showing the line profile post correction; and the pentagons correspond to the profile before correction of the decay observation with the stars showing the profile post correction.
As with Figure~\ref{fig:2017_out}(b), the line profiles in Figure~\ref{fig:2014}(b) retain their shape and have enhanced intensities across the lines.
The pre-flare corrected spectrum is the one that has changed the most with a noticeable increase in intensity towards the line core.
This is a result of the ``smearing'' of light mentioned in Section~\ref{sec:2017}.
Again, the Doppler shifts and intensity-averaged wavelengths are approximately conserved.
The line profiles from the rise-phase observations in Figure~\ref{fig:2014}(c) show a different story.
The shape of the line profile is not too dissimilar from the line profile before correction but the intensity values are estimated at around 2x higher.
This could be in part due to the ``smearing'' of the flare ribbon emission before correction but also due to over-estimation of the bright features, which is discussed more in Section~\ref{sec:conc}.
This correction is outside of the range of the error bars of the estimate and a more robust approach to error calculation for this model may be needed and is discussed further in Section~\ref{sec:conc}.
The profile in the decay phase is corrected in a similar manner to the profile in the peak of SOL20170906T12:02 (Section~\ref{sec:2017}), in that the shape develops the typical two-horn profile of H$\alpha$ in flare ribbons.

The sum of the DNs across the field-of-view at each wavelength for each of the cases before and after correction is also calculated as in Section~\ref{sec:2017}.
The results are shown in Table~\ref{tab:2014}.
These corrections show the same behaviour as for the AR12673 data with a decrease in the number of DNs by a few percent.
This implies that a further DN conservation metric could be implemented in Equation~\ref{eq:totalL} to offset this under-estimate.

The boxes in Figure~\ref{fig:2014}(d)--(o) reference the subfields shown in Figure~\ref{fig:2014_zoom}.
Again, this is to illustrate how well our model recovers small-scale features across the varied field-of-view. 
Figure~\ref{fig:2014_zoom}(a)--(d) show the northeasterly part of AR12157 for the pre-flare both in the red wing -- panels (a) \& (b) -- and line core -- panels (c) \& (d) -- of H$\alpha$.
Figure~\ref{fig:2014_zoom}(e)--(h) shows part of the sunspot umbra/penumbra in the previous format during the rise of the soft X-ray curve of the flare, and Figure~\ref{fig:2014_zoom}(i)--(l) part of the sunspot penumbra during the decay of the flare in the same format.
This figure, again, is for illustrative purposes and shows the good recovery of small-scale feature particularly in the line core. 
% lf{There are some slightly weird small-scale patterns appearing and I doubt all of them are real. eg panel b where it looks like the ridged pattern visible in panel a (probably an instrumental effect) has been enhanced and then repeated across the field.}

\begin{table}
\centering
\caption{The percentage changes in the total number of DNs between the non-corrected and corrected images using the same conventions as Table~\ref{tab:2017}.}
\label{tab:2014}
\begin{tabular}{cccc}
\toprule
 $\Delta \lambda$ [\AA{}] &  Pre-flare [\%] &  Rise of M1.1 [\%] &  M1.1 Decay [\%] \\
\midrule
                    -1.40 &           -4.38 &              -4.48 &            -4.27 \\
                    -1.20 &           -4.53 &              -4.66 &            -4.46 \\
                    -1.00 &           -4.54 &              -4.55 &            -4.63 \\
                    -0.80 &           -5.16 &              -5.13 &            -4.95 \\
                    -0.60 &           -5.65 &              -6.39 &            -5.91 \\
                    -0.40 &           -8.39 &              -8.72 &            -7.21 \\
                    -0.20 &          -10.39 &             -11.87 &            -9.64 \\
                     0.00 &          -11.88 &             -10.84 &           -10.87 \\
                     0.20 &          -10.78 &             -10.86 &           -11.56 \\
                     0.40 &           -8.85 &              -8.90 &            -8.42 \\
                     0.60 &           -6.29 &              -6.56 &            -6.17 \\
                     0.80 &           -5.30 &              -5.23 &            -5.57 \\
                     1.00 &           -4.77 &              -4.62 &            -4.36 \\
                     1.20 &           -4.59 &              -4.37 &            -4.27 \\
                     1.40 &           -4.09 &              -4.60 &            -4.42 \\
\bottomrule
\end{tabular}
\end{table}

\section{Discussion and Conclusions}
\label{sec:conc}
We have presented a new method for seeing correction of intensity (Stokes I) images in ground-based solar flare observations. This method can be adapted to other problems after generation of the training set.
In this method, a neural network is trained to learn to correct for synthetic seeing, generated by a mathematical model, which is applied to data observed in good seeing. This network is then applied to real data taken in bad seeing.
We found that the network performs best when the effects of seeing are minimal, as expected.
When seeing is worse, the network is still good at recovering large-scale features in the images (see bottom row in Figure~\ref{fig:2017_out}) but, on small scales, the reconstruction is perceivably less accurate (see Figure~\ref{fig:2017_out_zoom}(j,l)).
Moreover, when the seeing is worse, the network seems to over-compensate on the small scales introducing features that are not necessarily physical (again, see Figures~\ref{fig:2017_out_zoom}(j,l) \&~\ref{fig:2014_zoom}(l)).
On the other hand, the over-compensation may not be due to the bad seeing entirely as a small instrumental blemish can be seen in Figure~\ref{fig:2017_out_zoom}(k,l) just below the $y ~=~ 7.5^{\prime 
\prime}$ line at around $x ~=~ 12.5^{\prime \prime}$.
This takes the form of a Moir\'{e} pattern that may be introduced during the observation or the calibration of the data.
Further examples of this pattern appearing can be seen on larger scales in Figure~\ref{fig:2014}(h,l).
These patterns may cause inaccuracies in the reconstruction by the network.

An estimate of the error of the network was made by taking the final trained model and applying it to the training and validation sets combined and calculating the mean of the calculated losses by Equation~\ref{eq:totalL}.
This is a rather \emph{ad hoc} error which we wish to improve in the future using the method proposed in \citet{lowe_point-wise_1999} of training a network with an additional input that is the variance of the estimate that the network generates.
This will add a robustness to our error calculation and deliver a network capable of providing corrections and their confidence intervals.
This may help to solve the problem of the underestimation of DNs as discussed in Sections~\ref{sec:2017}~\&~\ref{sec:2014} as the discrepancy in DNs may be encapsulated by the confidence interval.

Furthermore, the change in DNs was considered for the training dataset which yielded interesting findings: the discrepancy in the number of DNs for the training set is always small (<1\%) and is both positive and negative (meaning that there is sometimes an overestimation and sometimes an underestimation).
This contrasts the examples we study in Sections~\ref{sec:2017}~\&~\ref{sec:2014}.
This means that the discrepancy in photons may just be apparent in the examples that we have chosen but it could also be systematic of the model we have trained.
In case of the latter, we propose, in the future, to have an added term to Equation~\ref{eq:totalL} in the optimisation of our system which minimises the difference in the total number of DNs between the ground truth and the corrected frames.

All in all, the model that has been trained produces nice corrections on spectroscopic images that would otherwise be plagued with bad seeing.
This allows us to study these flare events at higher time resolution more confidently as the geometry of the ribbons and their intensities have been corrected for bad seeing.
We note that our model only perform seeing correction for Stokes I and in the case of having full spectropolarimetric imaging it is hypothesised that the seeing in Stokes Q, U, and V can be corrected for using the method in \citet{diaz_baso_solar_2019}.

The code used here is available at \url{https://github.com/rhero12/Shaun2}.
The code also includes a model trained for corrections to Ca~\textsc{ii} 8542\AA{} data where seeing on average will be worse due to its longer wavelength.

\section*{Acknowledgements}

J.A.A. acknowledges support from `ScotDIST' doctoral training centre supported by grant ST/R504750/1 from the United Kingdrom Research and Innovation's (UKRI) Science and Technology Facilities Council (STFC).
L.F. acknowledges support from STFC grants ST/P000533/1 and ST/T000422/1.
The authors thank A. Reid \& M. Mathioudakis for providing the calibrated and prepped SST/CRISP data for AR12673.
J.A.A. also thanks C.M.J. Osborne for useful discussions and acronym wizardry.
J.A.A. further thanks P.J.A. Sim\~{o}es for useful discussions.
The authors thank the reviewer for their review and helpful comments.

%%%%%%%%%%%%%%%%%%%%%%%%%%%%%%%%%%%%%%%%%%%%%%%%%%



\bibliographystyle{mnras}
\bibliography{shaun} % if your bibtex file is called example.bib

% \begin{thebibliography}{99}
% \bibitem[\protect\citeauthoryear{Author}{2012}]{Author2012}
% Author A.~N., 2013, Journal of Improbable Astronomy, 1, 1
% \bibitem[\protect\citeauthoryear{Others}{2013}]{Others2013}
% Others S., 2012, Journal of Interesting Stuff, 17, 198
% \end{thebibliography}

%%%%%%%%%%%%%%%%%%%%%%%%%%%%%%%%%%%%%%%%%%%%%%%%%%

%%%%%%%%%%%%%%%%% APPENDICES %%%%%%%%%%%%%%%%%%%%%

\appendix
%%%%%%%%%%%%%%%%%%%%%%%%%%%%%%%%%%%%%%%%%%%%%%%%%%

% Don't change these lines
\bsp	% typesetting comment
\label{lastpage}
\end{document}